%% file: chmpg.tex
\documentclass[english,a4paper,twoside,aps,preprint,superscriptaddress]{revtex4}
\usepackage[T1]{fontenc}
\usepackage[latin9]{inputenc}
\usepackage{verbatim}
\usepackage{amsmath}
\usepackage{graphicx}

\makeatletter

\providecommand{\tabularnewline}{\\}

\makeatletter

\makeatletter



\makeatother

\makeatother

\usepackage{babel}
\makeatother

\begin{document}

\preprint{KIAS-P07092}

\title{Constraining Charming Penguins in Charmless $B$ Decays}

\author{Yue-Liang Wu$^{1}$, Yu-Feng Zhou$^{2}$ and Ci Zhuang}

\affiliation{Kavli Institute for Theoretical Physics China, Institute of Theoretical
Physics\\
 Chinese Academy of Sciences, Beijing, 100080, China \\
 $^{2}$Korea Institute for Advanced Study, Seoul 130-722, Korea }

\begin{abstract}
We discuss the correlations of charming penguin contributions to $B\to\pi\pi$,
$\pi K$ and $KK$ using approximate flavor $SU(3)$ symmetry. Strong
constraints are found from the direct CP asymmetries especially in
$\pi K$ modes. We make a global fit to the latest data and find that
only a small charming penguin is allowed, and the size of color-suppressed
tree amplitude ($C$) relative to tree amplitudes $(T)$ remains large
$C/T\simeq0.6$, which disfavors the possibility of a large charming
penguin as an explanation for the $\pi\pi$ puzzle. We show that a
small charming penguin can still have sizable effect in the time-dependence
CP asymmetries in $KK$ mode.
\end{abstract}
\maketitle

\section{Introduction}

With the successful running of the two $B$ factories, the $B$ physics
has entered a precision era. Although the current data of hadronic
$B$ decays has show an overall agreement with the Standard Model
(SM), there are a number of modes with unexpected decay rates and
CP asymmetries, which are often referred to as puzzles. One of the
puzzles in $\pi\pi$ modes is a large averaged branching ratio of
$\pi^{0}\pi^{0}$ relative to $\pi^{+}\pi^{-}$, the current data
read \cite{HFAG} \begin{align}
R_{\pi\pi} & =\frac{2Br(\pi^{0}\pi^{0})}{Br(\pi^{+}\pi^{-})}=0.51\pm0.08\label{eq:Rpipi}\end{align}
 which is significantly larger than theoretical estimations. Another
one in $\pi K$ is the difference in two direct CP asymmetries \begin{eqnarray}
A_{cp}(\pi^{+}K^{-})=-0.097\pm0.012 & , & A_{CP}(\pi^{0}K^{-})=0.050\pm0.025,\end{eqnarray}
 Both of the puzzles require a large color-suppressed tree amplitudes
in flavor SU(3) topology, which is difficult to obtain from
short-distance contributions. So far a satisfactory explanation is
not yet available. There are other potential puzzles regarding the
branching ratios and time-dependent CP asymmetries in $\pi K$ mode
which are relevant to the possibility of new physics. In the present
work, we focus on the former ones which are more relevant to the
hadronic dynamics.

It was emphasized in literature that long-distance Final State Interactions
(FSIs) may play important roles in these modes \cite{Wolfenstein:1990ks,Donoghue:1996hz,Suzuki:1999uc},
such as the inelastic rescattering channel $B\to DD_{(s)}\to\pi\pi(K),KK$
at meson level. Topologically they are equivalent to the charm-quark
loops in the contractions of local operators $Q_{1}^{c}=(\bar{d}c)_{V-A}(\bar{c}b)_{V-A}$
and $Q_{2}^{c}=(\bar{d}_{\alpha}c_{\beta})_{V-A}(\bar{c}_{\beta}b_{\alpha})_{V-A}$
at quark level which are referred to as charming penguins\cite{Ciuchini:1997hb,Isola:2001ar,Isola:2001bn}.
Experimentally $Br(B\to D^{+}D_{(s)}^{-})=[1.9\pm0.6(65\pm21)]\times10^{-4}$\cite{Yao:2006px}
are about 40(300) times larger than that of $B\to\pi^{+}\pi^{-}(K^{-})$.
As a consequence, a tiny OZI violating $DD\to\pi\pi$ mixing may lead
to significant changes in branching ratios and CP asymmetries in $\pi\pi$$(\pi K)$
modes \cite{Smith:2003iu,Barshay:2004hb,Barshay:2004ra}. A large charming penguin
with an appropriate strong phase may simultaneously suppress $Br(\pi^{+}\pi^{-})$
while enhance $Br(\pi^{0}\pi^{0})$, thus providing a solution to
the $\pi\pi$ puzzle.

The effects of charming loop have been discussed at both quark level
and meson level. The situation is not yet conclusive. Estimations
based on pQCD\cite{Li:2006cva} and QCD sum rules \cite{Khodjamirian:2003eq}
favor a small size. While in the framework of Soft Collinear Effective
Theory (SCET) the charming penguin could be large, depending on the
jet function\cite{Bauer:2004tj}. The meson level calculations using
effective Lagrangian for mesons favor a large charming penguin comparable
to QCD penguin in $\pi K$ \cite{Kamal:1998kk,Kamal:1999rn,Cheng:2004ru,Atwood:1998iw}.
But the patterns in $\pi\pi$ data can not be well reproduced \cite{Cheng:2004ru}.

Note that a global analysis using approximate flavor SU(3) symmetry
for all the $\pi\pi$, $\pi K$ and $KK$ modes may provide a powerful
constraint on charming penguins. This is because the presence of
charming penguin not only modifies the individual decay amplitude
but also changes the correlations among them, which has not been
enough emphasized in previous analysis. The correlations are of
particular importance in distinguishing charming penguin from QCD
penguin. First, although the two type of amplitudes always appear
together, in $\Delta S=1$ modes they are both nearly real, but in
$\Delta S=0$ modes they differ by a phase angle $\beta$ of the
unitarity triangle. The correlations in the predictions of direct CP
asymmetries are changed. Second, the $\Delta S=1$, $B\to\pi K$ modes
are penguin-dominant, which constrain the absolute size of charming
penguin together with QCD penguin, while the $\Delta S=0$,
$B\to\pi\pi$ modes are tree-dominant and more sensitive to the
tree-penguin interference. A strong constraint comes when they are
combined together. Finally, the $\Delta S=0$, $B\to KK$ modes
provides a testing ground for the charming penguin. In the SM, the
time-dependent $CP$ asymmetry $S(K_{S}K_{S})$ is nearly zero because
only QCD penguin contributes. The presence of charming penguin
provides an additional amplitude with different weak and strong
phases. Thus a significant deviation from zero is possible.

The present work is organized as follows. In section \ref{sec:charming-penguin},
we discuss the nontrivial correlations caused by long-distance charming
penguin using the QCD factorization results for short-distance contributions.
In section \ref{sec:Constrining}, we make a largely model-independent
global determination for the charming penguin using the latest data.
The results show that a small charming penguin is favored, which can
not play any significant role in resolving the $\pi\pi$ puzzle. But
it can still significantly affect the prediction for $S(K_{S}K_{S})$.
Some remarks and conclusions are in section \ref{sec:conclusions}.

\section{\label{sec:charming-penguin}charming penguin contribution to individual
modes}

The simplest way to see the correlated contributions from charming
penguins in different modes is to fix other hadronic amplitudes to
their theoretical values. To this end, we decompose the whole decay
amplitudes ($\mathcal{A}$) into short-distance ($\mathcal{A}_{SD}$)
and long-distance ($\mathcal{A}_{SL}$) part $\mathcal{A}=\mathcal{A}_{SD}+\mathcal{A}_{LD}$,
and take the short-distance part from theoretical calculations. The
long-distance part is assumed to be dominated by charming penguins.
The decay amplitudes are related to the observable of decay branching
ratio and direct CP asymmetry as follows \begin{align}
Br & =\frac{p_{c}\tau_{B}}{16\pi m_{B}^{2}}(|\mathcal{A}|^{2}+|\bar{\mathcal{A}}|^{2}),\quad a_{cp}=\frac{|\bar{\mathcal{A}}|^{2}-|\mathcal{A}|^{2}}{|\bar{\mathcal{A}}|^{2}+|\mathcal{A}|^{2}},\label{eq:}\end{align}
 where $p_{c}$ is the momentum of final state meson in the $B$ meson
rest frame and $\tau_{B}=1.530(1.638)\times10^{-12}s$ \cite{Yao:2006px}
is the neutral (charged) $B$ meson life-time. The time dependent
CP asymmetry is \begin{align}
a_{cp}(t) & =\frac{\Gamma(\bar{B}^{0}\to f_{CP})-\Gamma(B^{0}\to f_{CP})}{\Gamma(\bar{B}^{0}\to f_{CP})+\Gamma(B^{0}\to f_{CP})}\nonumber \\
 & =S\cdot\sin(\Delta m_{B}\cdot t)-C\cdot\cos(\Delta m_{B}\cdot t).\label{eq:}\end{align}
 The definition of quantities $S$ and $C$ are given by \begin{equation}
S=\mbox{Im}\left(\frac{q}{p}\frac{\bar{\mathcal{A}}}{\mathcal{A}}\right),\quad C=\frac{|\mathcal{A}|^{2}-|\bar{\mathcal{A}}|^{2}}{|\bar{\mathcal{A}}|^{2}+|\mathcal{A}|^{2}}=-a_{cp},\label{eq:S
and C}\end{equation}
 where $(q/p)=e^{-2i\beta}$ in the SM with $\beta$ one of the anlges
of the unitarity triangle (UT). In what follows we take the CKM matrix
elements $V_{ub}$ and $V_{cb}$ from the global CKM fits\cite{Charles:2004jd}
\begin{eqnarray}
V_{ub}=(3.57\pm0.17)\times10^{-3},\quad V_{cb}=0.0405_{-0.0029}^{+0.0032}.\label{eq:Vub-Vcb-input}\end{eqnarray}
 To fix the profile of the UT we also use the best fitted value of
\cite{Charles:2004jd} \begin{eqnarray}
\gamma & = & 1.170_{-0.079}^{+0.048},\label{eq:gamma-input}\end{eqnarray}
 which corresponds to a best fitted $\beta=0.379\pm0.017$.

Recently the theoretical calculations for hadronic matrix elements
have been improved to next to leading $\alpha_{s}^{2}$ order (NLO)
in the framework of QCD factorization for spectator scaterings
\cite{Beneke:2005vv,Beneke:2006mk,Kivel:2006xc,Pilipp:2007mg} and in perturbative QCD (pQCD)
\cite{Li:2005kt,Li:2006cva}. In QCD factorization approach, the hard
spectator scattering effects can lift a cancellation between leading
term and vertex corrections, resulting in a significant enhancement
in the effective coefficient $\alpha_{2}(\pi\pi)$ by a factor of
$\sim3$ and improve the agreement with the data. Nevertheless
generating a large enough spectator scattering effects still require
tuning of input parameters and the latest calculation still favor a
$Br(\pi^{0}\pi^{0})$ lower than the current data \cite{Beneke:2006mk}. 
It remains to be seen if there is futher enhancement from NNLO calculations
\cite{Bell:2006tz,Bell:2007tv}.
Note that in the pQCD approach, although the NLO results improve the predictions for
the direct CP asymmetries in $\pi K$ modes, there is no significant
enhancement found in $\pi^{0}\pi^{0}$.

The whole charmless $B$ decay amplitudes can be described by a set
of flavor topological quark flavor flow diagrams \cite{Gronau:1995hn,Gronau:1998fn,Gronau:2000pk,Gronau:2002gj,Chau:1990ay}.
In this approach the decay amplitudes are expressed in terms of diagrams
such as tree ($T$), color-suppressed tree ($C$), QCD penguin ($P$
and $P_{tu}$), electroweak penguin ($P_{EW}$,), color-suppressed
electroweak penguin ($P_{EW}^{C}$) etc. In the presence of charming
penguin ( denoted by $D$ for $D\bar{D}$ intermediate states ), the decay amplitudes for $\pi\pi$ modes are given
by \begin{alignat}{1}
-\mathcal{\bar{\mathcal{A}}}(\pi^{+}\pi^{-}) & =\lambda_{u}(T+E-P_{tu}-P_{A}-\frac{2}{3}P_{EW})-\lambda_{c}(P-D+P_{A}+\frac{2}{3}P_{EW}^{C}),\nonumber \\
-\mathcal{\bar{\mathcal{A}}}(\pi^{0}\pi^{0}) & =\frac{1}{\sqrt{2}}[\lambda_{u}(C-E+P_{tu}+P_{A}-P_{EW}-\frac{1}{3}P_{EW}^{C})+\lambda_{c}(P-D+P_{A}-P_{EW}-\frac{1}{3}P_{EW}^{C})],\nonumber \\
-\mathcal{\bar{\mathcal{A}}}(\pi^{0}\pi^{-}) & =\frac{1}{\sqrt{2}}[\lambda_{u}(T+C-P_{EW}-P_{EW}^{C})-\lambda_{c}(P_{EW}+P_{EW}^{C})].\label{eq:}\end{alignat}
 The CKM factors are defined as $\lambda_{q}^{(s)}=V_{qd(s)}^{*}V_{qb}$.
In general, the QCD penguin has three part $\lambda_{u}P_{u}+\lambda_{c}P_{c}+\lambda_{t}P_{t}$,
which is recombined as $P_{tu}\equiv P_{t}-P_{u}$ and $P\equiv P_{tc}\equiv P_{t}-P_{c}$.
The amplitudes $T,$ $C$ and $P$ etc can be calculated and the typical
values (in units of $10^{4}$eV) from QCD factorization are \begin{alignat}{1}
T & =A_{\pi\pi}a_{1,\pi\pi}\simeq0.89-0.02i,\nonumber \\
C & =A_{\pi\pi}a_{2,\pi\pi}\simeq0.24-0.02i,\nonumber \\
P_{tu} & =-A_{\pi\pi}(a_{4,\pi\pi}^{u}+r_{\chi}^{\pi}a_{6,\pi\pi}^{u})\simeq0.076+0.029i,\nonumber \\
P & =-A_{\pi\pi}(a_{4,\pi\pi}^{c}+r_{\chi}^{\pi}a_{6,\pi\pi}^{c})\simeq0.084+0.015i,\end{alignat}
 where $A_{\pi\pi}=G_{F}f_{\pi}F_{0}^{B\to\pi}(m_{B}^{2}-m_{\pi}^{2})/\sqrt{2}$.
The numerical values in the above expressions are in accordance with
the central values of NLO effective coefficients in QCD factorization
approach \cite{Beneke:2006mk} \begin{eqnarray}
\alpha_{1,\pi\pi} & = & 0.975_{-0.072}^{+0.034}+(-0.017_{-0.051}^{+0.022})i,\nonumber \\
\alpha_{2,\pi\pi} & = & 0.275_{-0.135}^{+0.228}+(-0.024_{-0.081}^{+0.115})i,\nonumber \\
\alpha_{4,\pi\pi}^{u} & = & -0.024_{-0.002}^{+0.004}+(-0.012_{-0.002}^{+0.003})i,\nonumber \\
\alpha_{4,\pi\pi}^{c} & = & -0.028_{-0.003}^{+0.005}+(-0.006_{-0.002}^{+0.003})i,\nonumber \\
r_{\chi}^{\pi}\alpha_{6,\pi\pi}^{u} & = & -0.060_{-0.017}^{+0.001}+(-0.020_{-0.006}^{+0.005})i,\nonumber \\
r_{\chi}^{\pi}\alpha_{6,\pi\pi}^{c} & = & -0.065_{-0.019}^{+0.012}+(-0.010_{-0.004}^{+0.004})i.\end{eqnarray}
 The short-distance calculations suggest a $t-$quark dominance in
QCD penguin such that $P_{tu}\simeq P$, and tiny annihilation type
diagrams $E,A$ and $P_{A}$ which are power suppressed.

\subsection{$\pi\pi$ modes}

We begin with a re-examination of $\pi\pi$ puzzle in the presence
of charming penguin $D$. In the limit of $T,C\gg P,D$, the ratio
$R_{\pi\pi}$ can be expanded as follows \begin{align}
R_{\pi\pi} & \simeq\frac{C^{2}}{T^{2}}\left[1+2\left(1-\omega\cos\gamma\right)\left(\frac{P}{T}\cos(\delta_{T}-\delta_{P})+\frac{P}{C}\cos(\delta_{C}-\delta_{P})\right)\right.\nonumber \\
 & +\left.2\omega r_{D}\left(\frac{P}{T}\cos(\delta_{T}-\delta_{D})+\frac{P}{C}\cos(\delta_{C}-\delta_{D})\right)\cos\gamma\right]\label{eq:}\end{align}
 where $\omega=|\lambda_{c}/\lambda_{u}|\simeq2.73$, and $r_{D}\equiv D/P$
is the size of charming penguin relative to QCD penguin. It is evident
that the charming penguin has opposite contributions to $\pi^{+}\pi^{-}$
and $\pi^{0}\pi^{0}$ modes. In order to enhance $R_{\pi\pi}$ one
needs $\cos(\delta_{C}-\delta_{D})>0$ and a large $r_{D}$. In Fig.
1 
we plot the ratio $R_{\pi\pi}$ as a function of $r_{D}$ with different
strong phases. In the numerical calculations we use the full expressions
for decay rates and CP asymmetries. %
\begin{figure}
\includegraphics[width=0.65\textwidth]{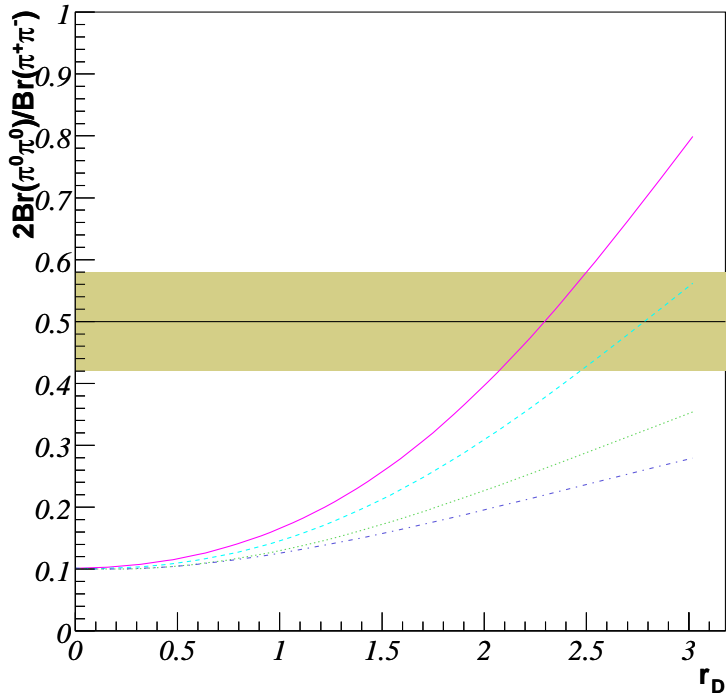}

\caption{\label{Rpipi} The ratio $R_{\pi\pi}$ as function of charming penguin.
Four curves corresponds to the strong phase $\delta_{D}=30^{\circ}$(solid),
$60^{\circ}$(dashed), $90^{\circ}$(dotted) and $120^{\circ}$(dot-dashed)
respectively. Other parameters are default in QCD factorization estimations. }

\end{figure}

It is shown in the figure that for a typically small strong phase
$\delta_{D}=30^{\circ}$, a large $2\leq r_{D}\leq2.5$ is needed to
meet data of $R_{\pi\pi}$. For large strong phase
$\delta_{D}>90^{\circ}$ , an even larger $r_{D}>3$ is required. This
confirms previous phenomenological studies in favor of large
charming penguin. The direct CP asymmetric measurements provide
different constraints. In the limit $T,C\gg P,D$, the direct CP
asymmetries are approximated by \begin{align}
a_{cp}(\pi^{+}\pi^{-}) & \simeq2\omega\frac{P}{T}\left(\sin(\delta_{T}-\delta_{P})-r_{D}\sin(\delta_{T}-\delta_{D})\right)\sin\gamma,\nonumber \\
a_{cp}(\pi^{0}\pi^{0}) & \simeq-2\omega\frac{P}{C}\left(\sin(\delta_{C}-\delta_{P})-r_{D}\sin(\delta_{C}-\delta_{D})\right)\sin\gamma.\label{eq:}\end{align}
 Similar to the $\pi\pi$ decay rates, the charming penguin contributions
to the direct CP asymmetries are again opposite. Since $\delta_{T}$
and $\delta_{C}$ are small, roughly speaking for $0\leq\delta_{D}\leq180^{\circ}$,
it enhances $a_{cp}(\pi^{+}\pi^{-})$ while suppresses $a_{cp}(\pi^{0}\pi^{0})$
to negative values. The numerical results are shown in Fig.2. 
\begin{figure}
\includegraphics[width=0.45\textwidth]{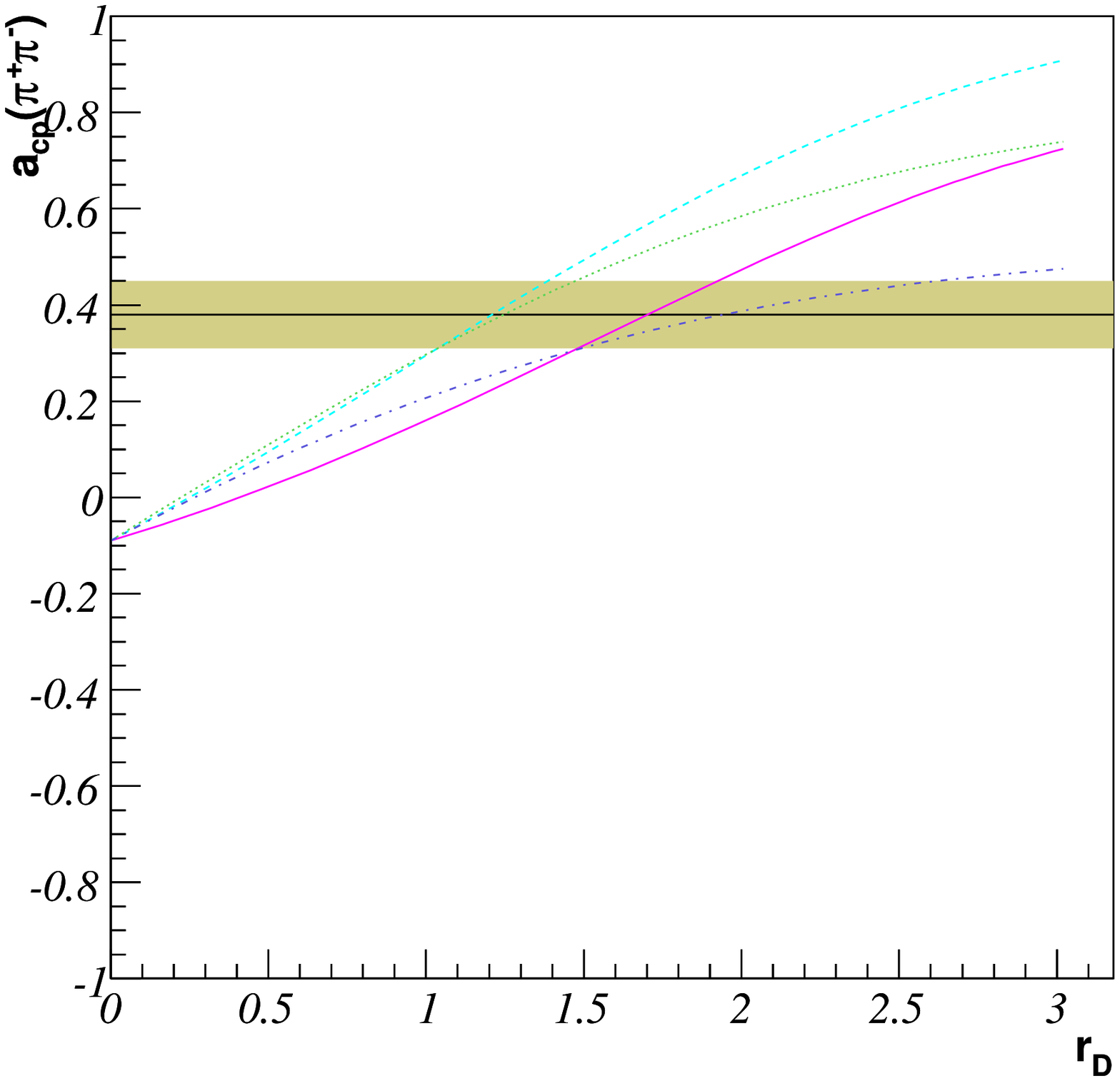} \includegraphics[width=0.45\textwidth]{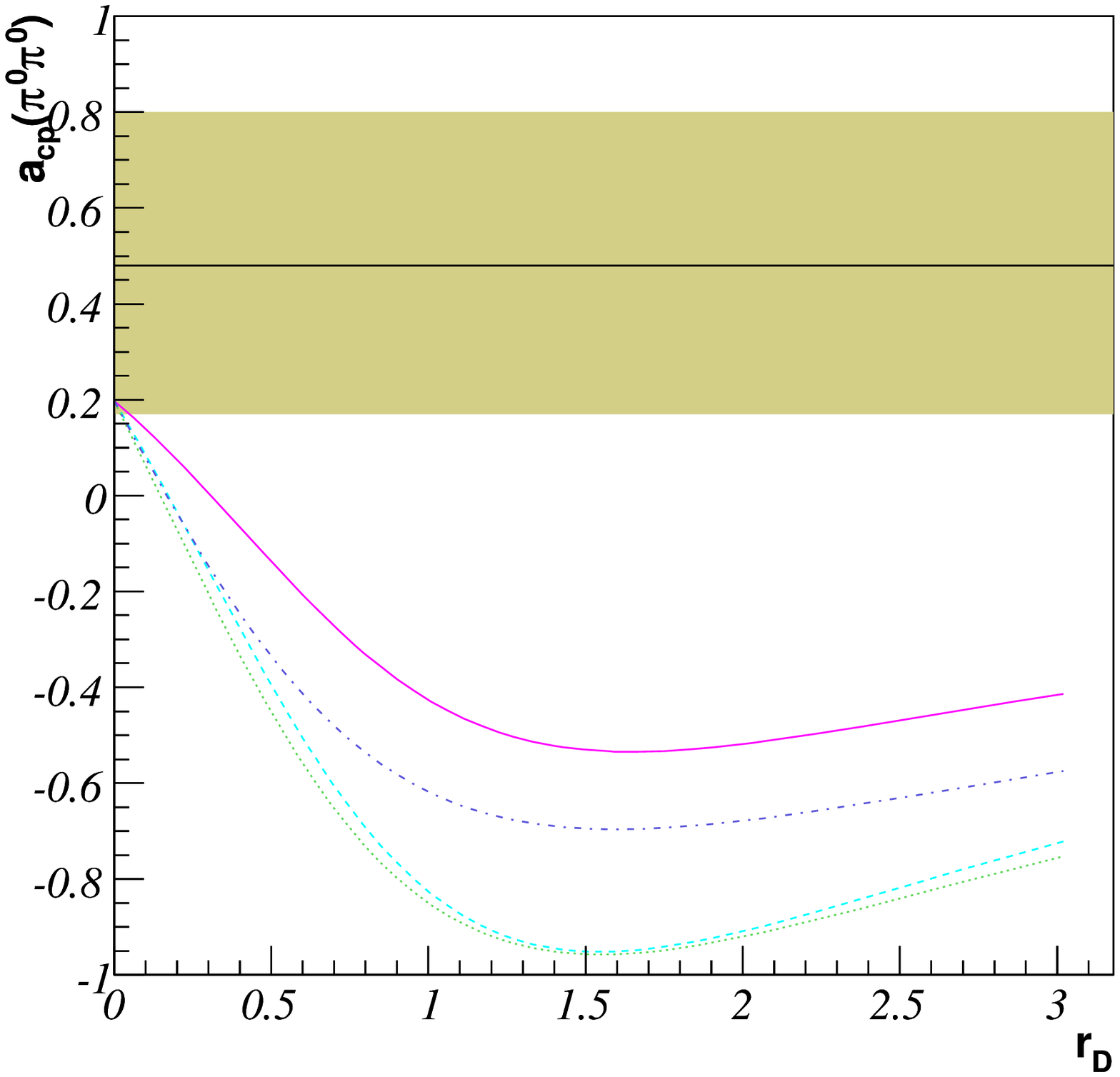}

\caption{\label{fig:cpPIPI}$a_{cp}(\pi^{+}\pi^{-})$ and $a_{cp}(\pi^{0}\pi^{0})$
as function of $D$. Four curves corresponds to the strong phase $\delta_{D}=30^{\circ}$(solid),
$60^{\circ}$(dashed), $90^{\circ}$(dotted) and $120^{\circ}$(dot-dashed)
respectively. Other parameters are default in QCD factorization estimations. }

\end{figure}

For small phase $\delta_{D}=30^{\circ}$, the current data of
$a_{cp}(\pi^{+}\pi^{-})$ restricts the size of $r_{D}$ to be
$1.5<r_{D}<2$. Note that there is still a significant difference
between two $B$ factories on the measurement of
$a_{cp}(\pi^{+}\pi^{-})$\cite{Ishino:2006if,Aubert:2007mj}
\begin{eqnarray}
a_{cp}(\pi^{+}\pi^{-}) & = & 0.21\pm0.09\pm0.02(Babar)\nonumber \\
 & = & 0.55\pm0.08\pm0.05(Belle)\end{eqnarray}
 The Babar measurement favors a smaller $a_{cp}(\pi^{+}\pi^{-})$
and the constraints on $r_{D}$ is stronger. Note that the constraints
on the size of $r_{D}$ and the strong phase $\delta_{D}$ from the
$a_{cp}(\pi^{0}\pi^{0})$ and $R_{\pi\pi}$ are opposite. The preliminary
data although with large uncertainty are in favor of a positive $a_{cp}(\pi^{0}\pi^{0})$,
which disfavor any large value of $r_{D}$ with $\delta_{D}$ in the
range $(0,\pi)$.

\begin{figure}
\includegraphics[width=0.45\textwidth]{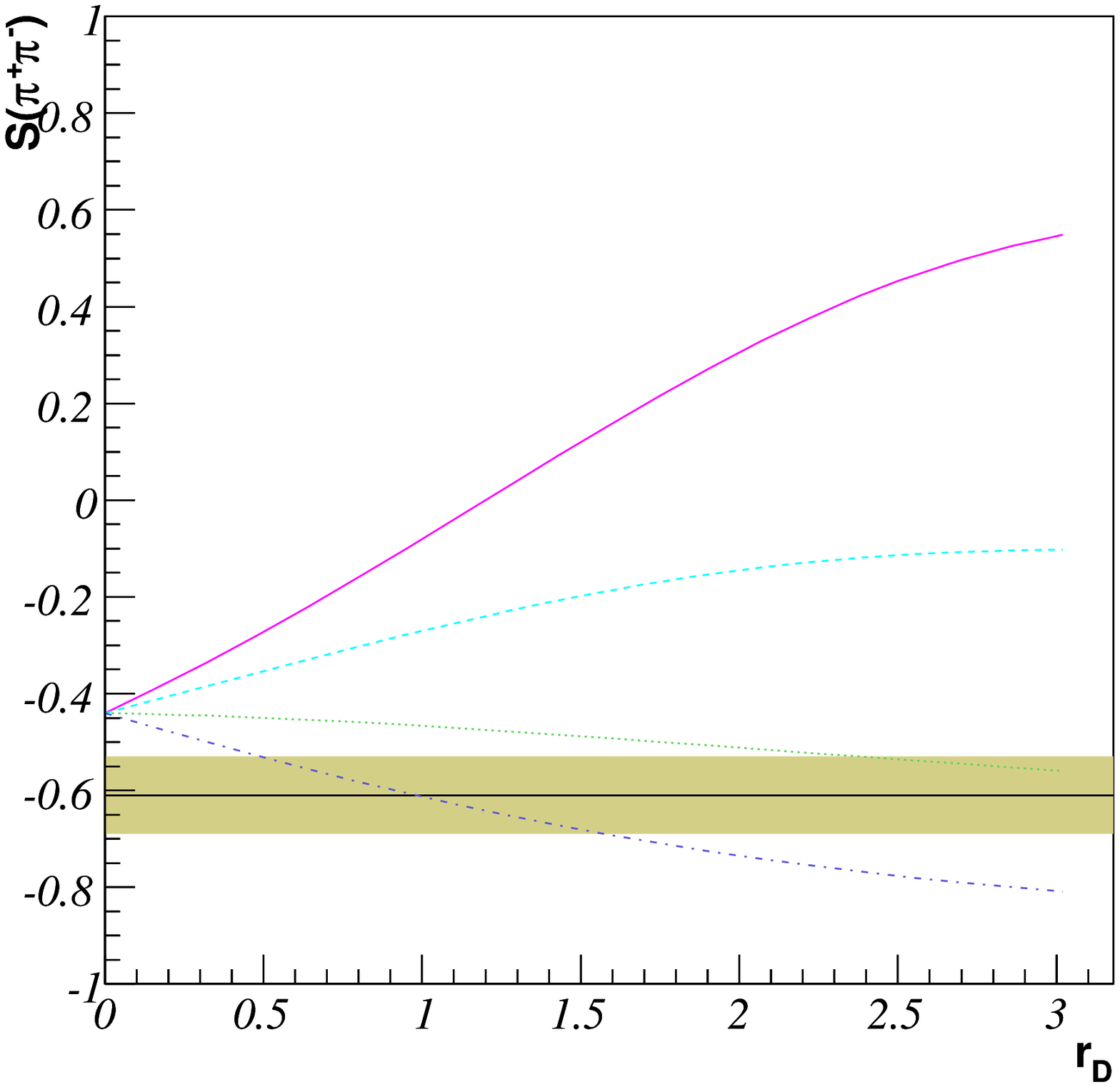} \includegraphics[width=0.45\textwidth]{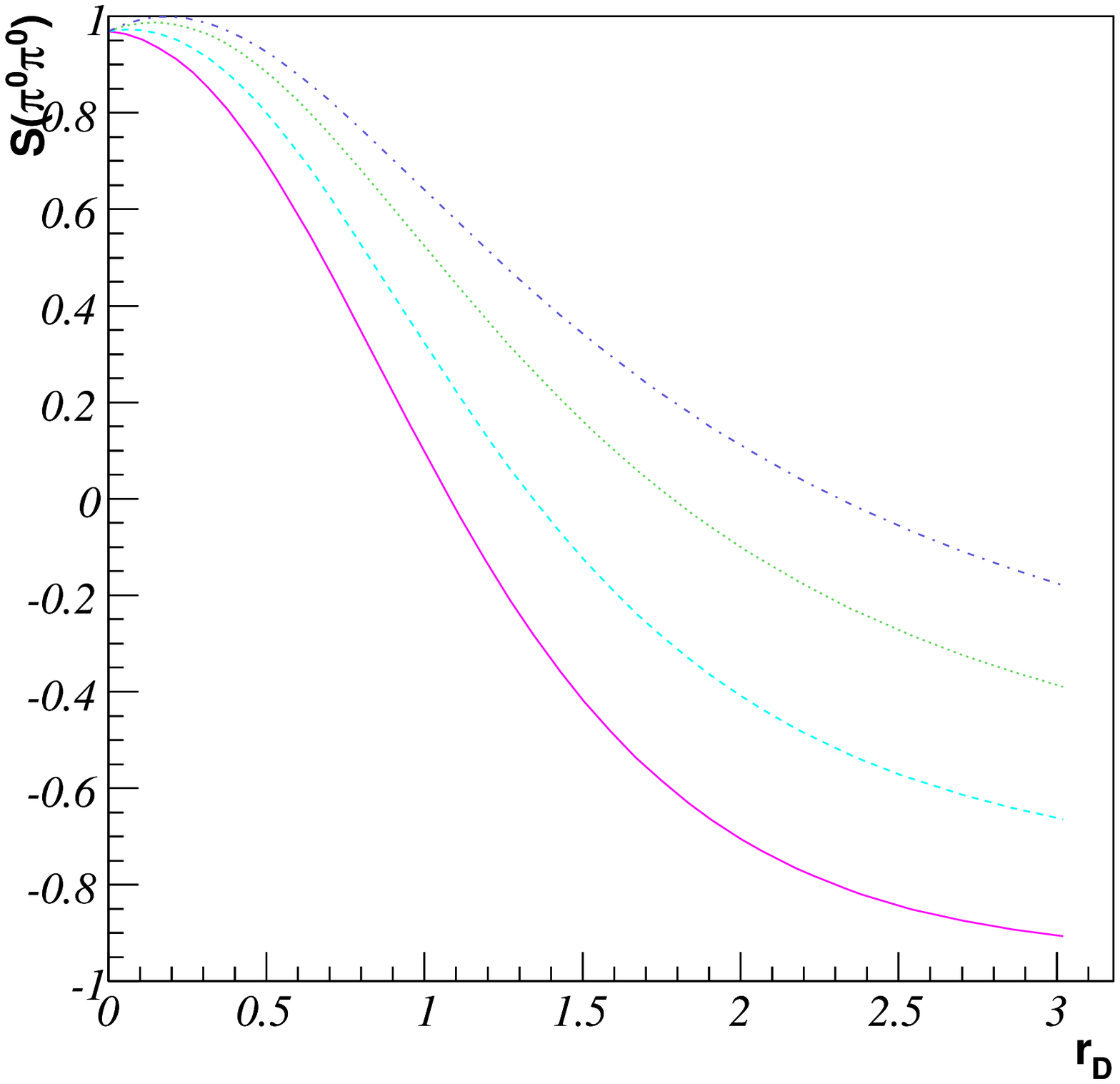}

\caption{\label{fig:sPIPI}$S(\pi^{+}\pi^{-})$ and $S(\pi^{0}\pi^{0})$ as
function of $r_{D}$. Four curves corresponds to the strong phase
$\delta_{D}=30^{\circ}$(solid), $60^{\circ}$(dashed), $90^{\circ}$(dotted)
and $120^{\circ}$(dot-dashed) respectively. Other parameters are
default in QCD factorization estimations. }

\end{figure}

A more significant $r_{D}$ dependence can be seen in the time-dependent
CP asymmetries which are approximated by \begin{align}
S(\pi^{+}\pi^{-}) & \simeq-\sin2(\beta+\gamma)+2\omega\frac{P}{T}\left(\cos(\delta_{P}-\delta_{T})-r_{D}\cos(\delta_{D}-\delta_{T})\right)\sin\gamma\cos2(\beta+\gamma),\nonumber \\
S(\pi^{0}\pi^{0}) & \simeq-\sin2(\beta+\gamma)-2\omega\frac{P}{C}\left(\cos(\delta_{P}-\delta_{C})-r_{D}\cos(\delta_{D}-\delta_{C})\right)\sin\gamma\cos2(\beta+\gamma).\end{align}
 Since the current global CKM fitting give a $\beta+\gamma$ close
to $\pi/2$, the leading term is suppressed for both $\pi^{+}\pi^{-}$
and $\pi^{0}\pi^{0}$, which significantly enhances the charming penguin
effects. As shown in Fig.\ref{fig:sPIPI}, for $\delta_{D}\le60^{\circ}$
the data of $S(\pi^{+}\pi^{-})$ exclude the possibility of charm
penguin since the short distance contribution is already above the
experiments. The data of $S(\pi^{+}\pi^{-})$ favors a larger strong
phase $\delta_{D}$. The charming penguin contribution can be either
positive and negative, depending on $\cos(\delta_{D}-\delta_{T})$.
Note that in $\pi^{0}\pi^{0}$ mode the charming penguin contribution
is proportional to $P/C$ much larger than that in $\pi^{+}\pi^{-}$
which is proportional to $P/T$. Thus $S(\pi^{0}\pi^{0})$ has the
strongest charming penguin dependence, which can be clearly seen from
Fig.\ref{fig:sPIPI}. For $\delta_{D}=30^{\circ}$ and $r_{D}=1$,
the value of $S(\pi^{0}\pi^{0})$ can be reduced to around zero. The
future precision measurement of $S(\pi^{0}\pi^{0})$ will provide
a very strong constraint on $r_{D}$. %

\subsection{$\pi K$ modes}

\begin{figure}
\includegraphics[width=0.45\textwidth]{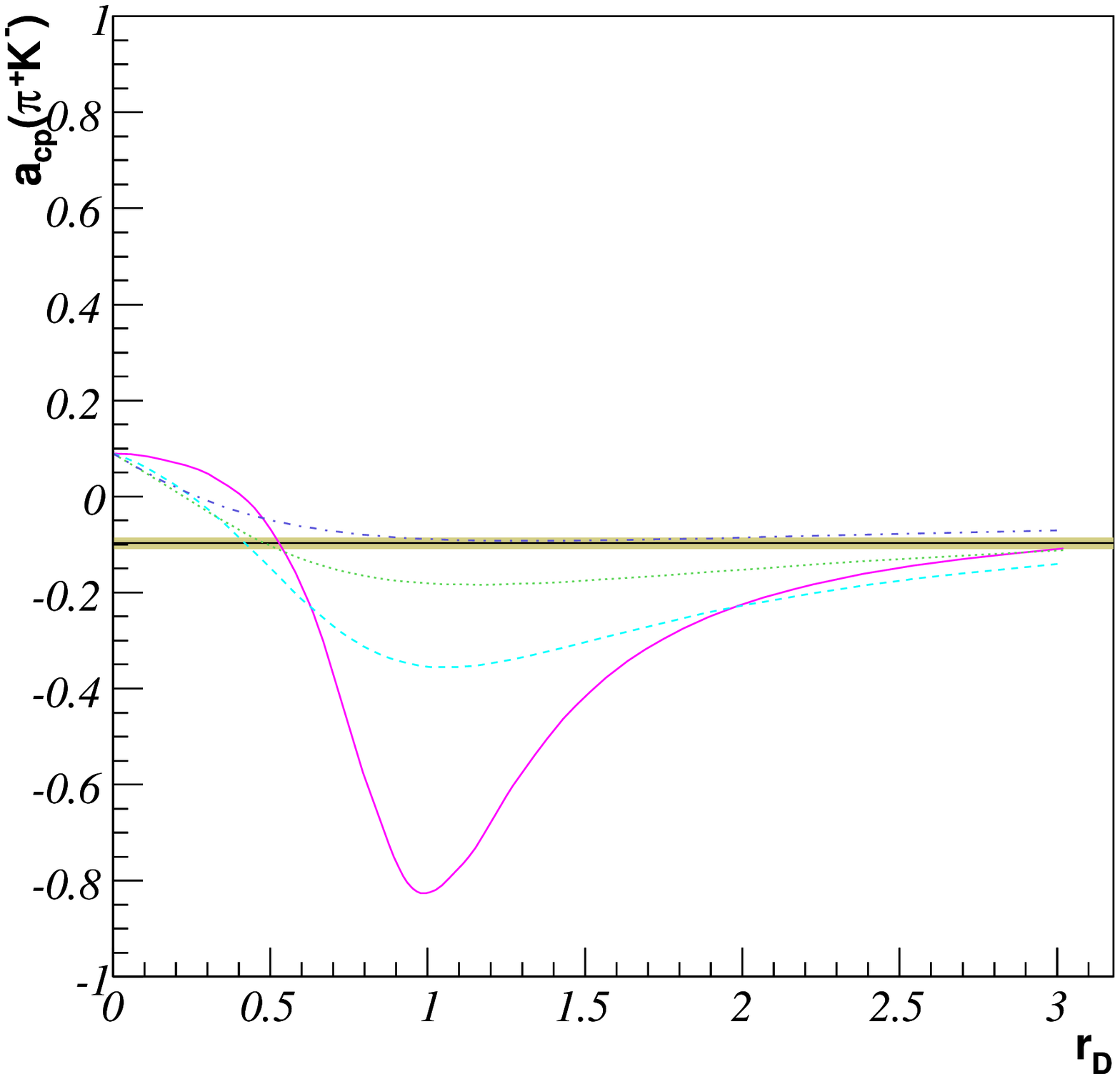}\includegraphics[width=0.45\textwidth]{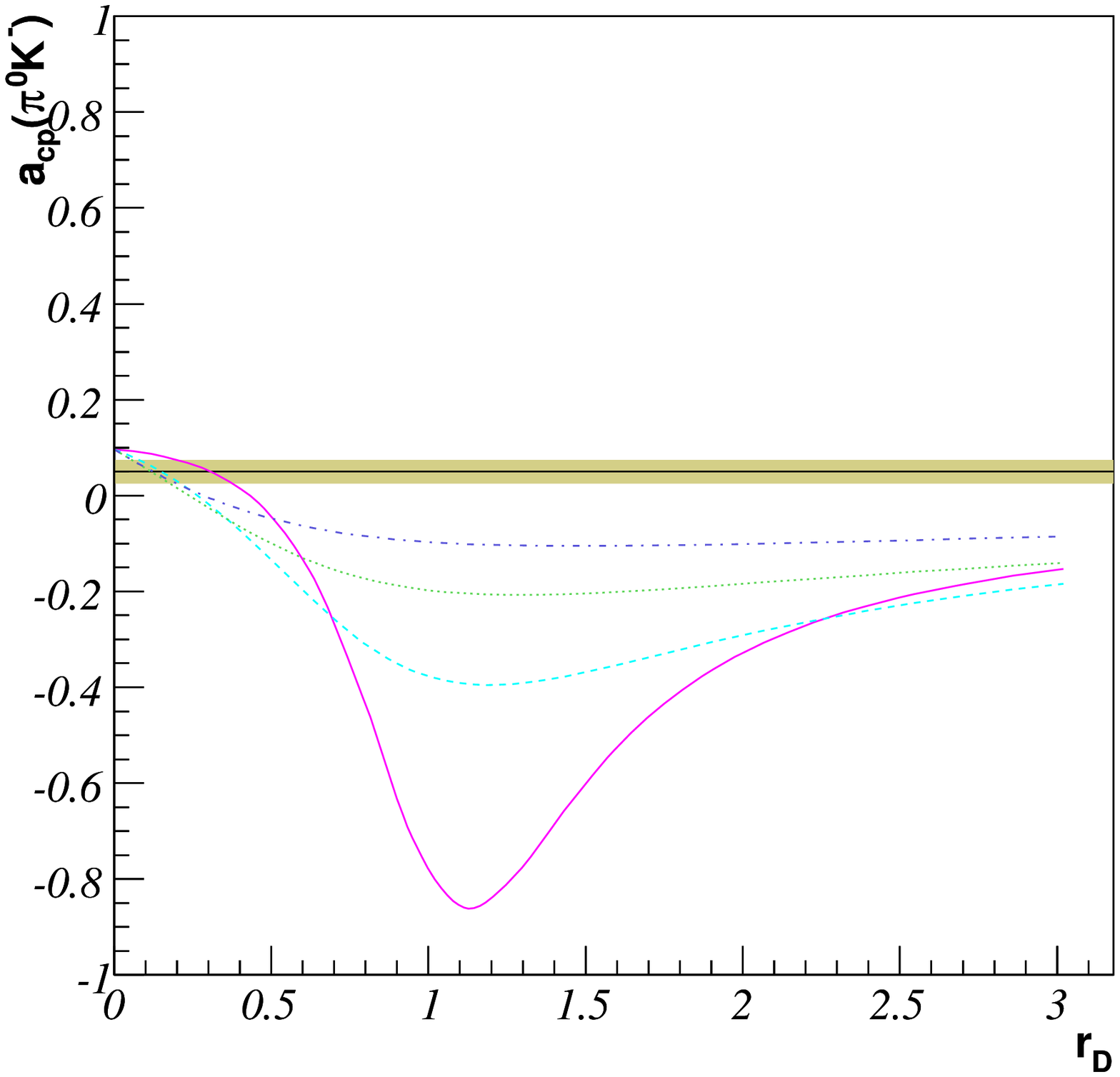}

\includegraphics[width=0.45\textwidth]{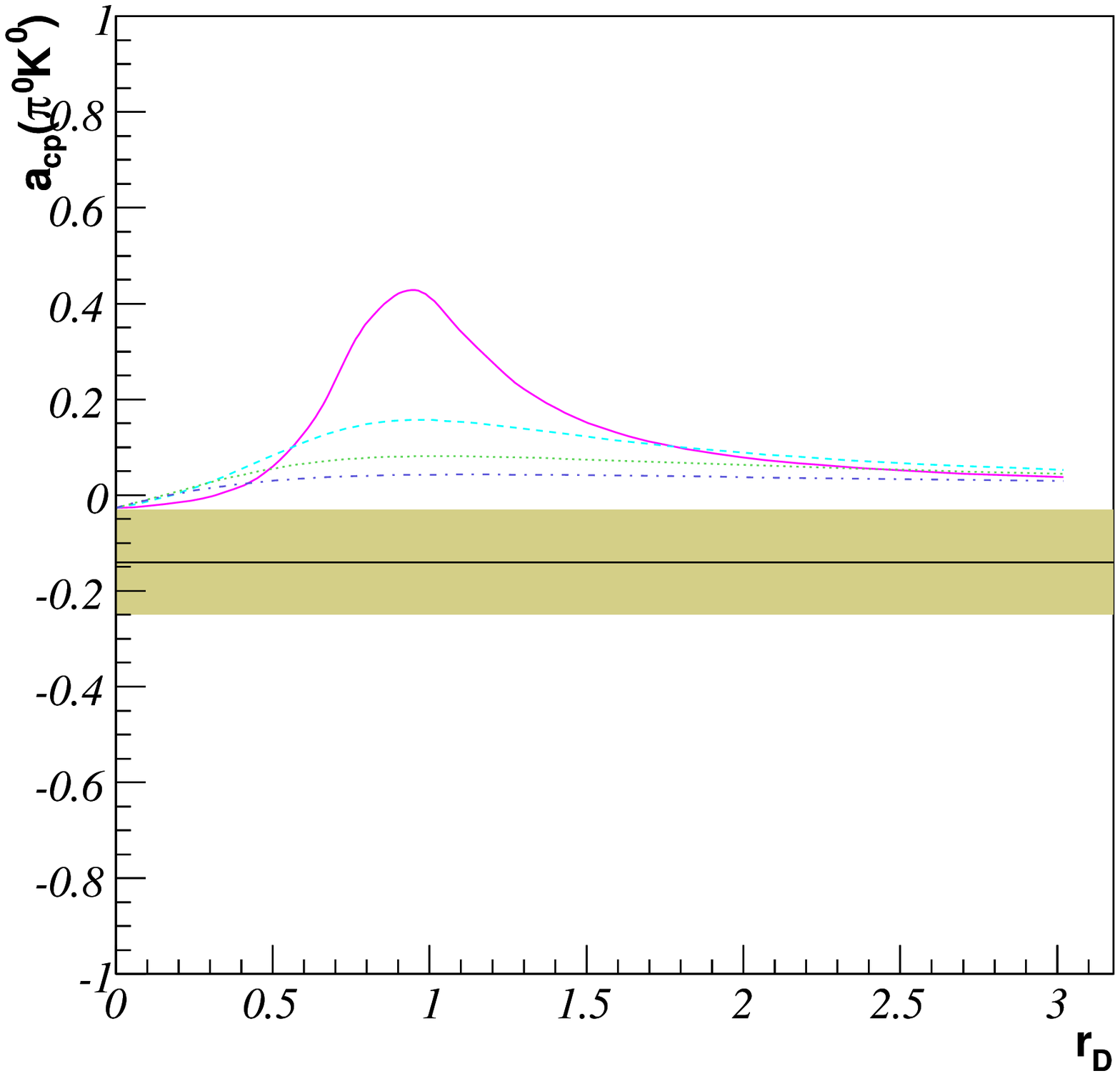}\includegraphics[width=0.45\textwidth]{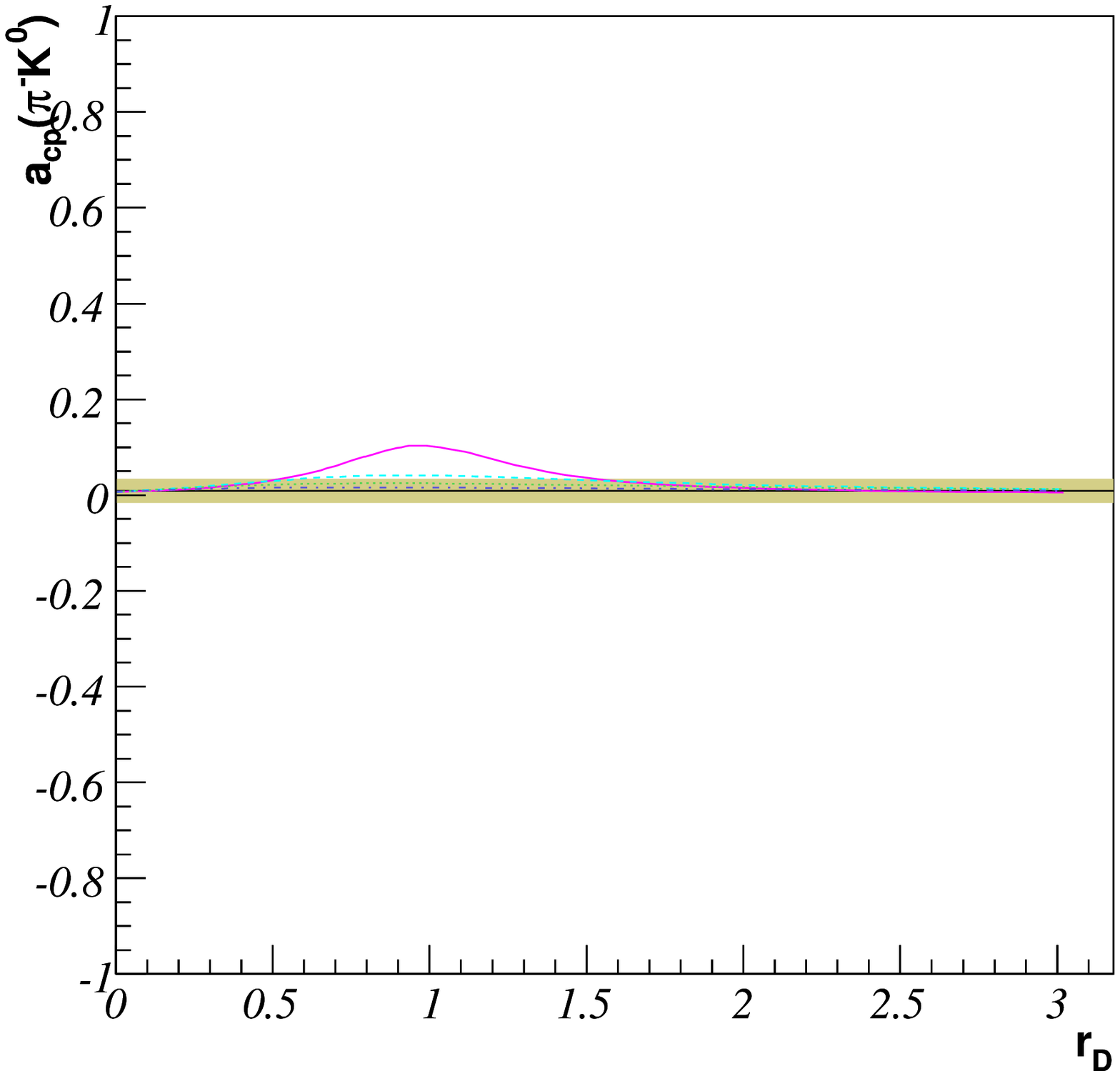}

\caption{\label{fig:cpPIK}$a_{cp}(\pi^{+}K^{-})$, $a_{cp}(\pi^{0}\bar{K}^{0})$,
$a_{cp}(\pi^{-}\bar{K}^{0})$ and $a_{cp}(\pi^{0}K^{-})$ as functions
of $r_{D}$. Four curves corresponds to the strong phase $\delta_{D}=30^{\circ}$(solid),
$60^{\circ}$(dashed), $90^{\circ}$(dotted) and $120^{\circ}$(dot-dashed)
respectively. Other parameters are default in QCD factorization estimations.}

\end{figure}

We process to discuss the $\pi K$ modes. The diagrammatic amplitudes
for $\pi K$ modes are given by \begin{alignat}{1}
-\mathcal{\bar{\mathcal{A}}}(\pi^{+}K^{-}) & =\lambda_{u}^{s}(T-P_{tu}-\frac{2}{3}P_{EW})-\lambda_{c}^{s}(P-D+\frac{2}{3}P_{EW}^{C})\nonumber \\
-\mathcal{\bar{\mathcal{A}}}(\pi^{0}\bar{K}^{0}) & =\frac{1}{\sqrt{2}}[\lambda_{u}^{s}(C+P_{tu}-P_{EW}-\frac{1}{3}P_{EW}^{C})+\lambda_{c}^{s}(P-D-P_{EW}-\frac{1}{3}P_{EW}^{C})]\nonumber \\
\mathcal{\bar{\mathcal{A}}}(\pi^{-}\bar{K}^{0}) & =\lambda_{u}^{s}(A-P_{tu}+\frac{1}{3}P_{EW}^{C})-\lambda_{c}^{s}(P-D-\frac{1}{3}P_{EW}^{C})\nonumber \\
-\mathcal{\bar{\mathcal{A}}}(\pi^{0}K^{-}) & =\frac{1}{\sqrt{2}}[\lambda_{u}(T+C+A-P_{tu}-P_{EW}-\frac{2}{3}P_{EW}^{C})-\lambda_{c}(P-D+P_{EW}+\frac{2}{3}P_{EW}^{C})]\label{eq:}\end{alignat}
 where the amplitudes $T,C$ and $P$ for $\pi K$ modes can be obtained
by replacing $A_{\pi\pi}$ into $A_{\pi K}$. The electroweak penguin
amplitude is calculated from the effective coefficients $a_{7,\pi K}$
and $a_{9,\pi K}$ \begin{alignat}{1}
P_{EW} & =\frac{3}{2}A_{\pi K}(a_{7,\pi K}^{c}-a_{9,\pi K}^{c})\simeq(1.317+0.015i)\times10^{-2},\nonumber \\
P_{EW,tu} & =\frac{3}{2}A_{\pi K}(a_{7,\pi K}^{c}-a_{9,\pi K}^{c})\simeq(1.315-0.014i)\times10^{-2}\end{alignat}
 which corresponds to the effective coefficients\cite{Beneke:2006mk}
\begin{eqnarray}
a_{7,\pi K}^{u}=0.058_{-0.017}^{+0.024}+(0.015_{-0.006}^{+0.010})i & , & a_{7,\pi K}^{c}=0.010_{-0.017}^{+0.011}+(0.000_{-0.006}^{+0.003})i,\nonumber \\
a_{9,\pi K}^{u}=-0.819_{-0.042}^{+0.080}+(0.029_{-0.023}^{+0.053})i & , & a_{9,\pi K}^{c}=-0.868_{-0.026}^{+0.058}+(0.015_{-0.018}^{+0.043})i.\end{eqnarray}
 One sees again that the $t-$quark dominance leads to $P_{EW}\simeq P_{EW,tu}$.
In what follows we neglect teh subleading color-suppressed diagram
$P_{EW}^{C}$. In $\pi K$ modes one can define the following two
ratios for neutral and charged modes \cite{HFAG} \begin{align}
R_{n}=\frac{Br(\pi^{+}K^{-})}{2Br(\pi^{0}\pi^{0})}=0.98\pm0.07 & ,\quad R_{c}=\frac{2Br(\pi^{0}K^{-})}{Br(\pi^{-}\bar{K}^{0})}=1.12\pm0.07.\end{align}
 The penguin dominance leads to an estimation of $R_{n}\approx R_{c}\approx1$.
Due to a cancelation in the subleading term, $R_{n}\simeq R_{c}$
holds to a high accuracy \cite{Wu:2006yj,Wu:2006ur}. Although
earlier data showed a small $R_{n}$, which is usually referred to as
$\pi K$ puzzle, the latest measurements show that this puzzle has
been significantly reduced. It is easy to see that the dominant
charming penguin contributions cancel out in both $R_{n}$ and
$R_{c}$ and the remaining parts are suppressed by a CKM factor
$\xi=|\lambda_{u}^{s}/\lambda_{c}^{s}|\simeq0.02$
\begin{align}
R_{n}\simeq R_{c} & \simeq1-2\xi\left[\frac{T}{P}\cos(\delta_{T}-\delta_{P})+\frac{C}{P}\cos(\delta_{C}-\delta_{P})-\frac{T}{P}r_{D}\cos(\delta_{T}-\delta_{D})\right.\nonumber \\
 & \left.-\frac{C}{P}r_{D}\cos(\delta_{C}-\delta_{D})\right]\cos\gamma+2\frac{P_{EW}}{P}\cos(\delta_{P_{EW}}-\delta_{P}).\label{eq:}\end{align}
 The previous global fits without charming penguin show a remarkable
agreement between theory and experiment in penguin amplitudes \cite{Wu:2005hi,Chiang:2006ih,Chiang:2007qh}.
The pure penguin mode such as $\pi^{-}\bar{K}^{0}$ constrain strongly
the absolute values of $|Pe^{i\delta_{P}}-De^{i\delta_{D}}|$. However,
the charming penguin can still be sizable when it carries large relative
strong phase. In this case, due to tiny but nonzero weak phase difference
between $D$ and $P$, unacceptably large CP asymmetries can be induced
when the charming penguin and QCD penguin are comparable in size.
The charming penguin contributions to the direct CP asymmetries in
the limit $D\ll P$ are given by \begin{align}
a_{CP}(\pi^{+}K^{-}) & \simeq-2\xi\frac{T}{P}\left[\sin(\delta_{T}-\delta_{P})-r_{D}\sin(\delta_{T}-\delta_{D})\right]\sin\gamma,\nonumber \\
a_{CP}(\pi^{0}\bar{K}^{0}) & \simeq2\xi\frac{C}{P}\left[\sin(\delta_{C}-\delta_{P})-r_{D}\sin(\delta_{C}-\delta_{D})\right]\sin\gamma,\nonumber \\
a_{CP}(\pi^{-}\bar{K}^{0}) & \simeq2\xi r_{D}\sin(\delta_{D}-\delta_{P})\sin\gamma,\nonumber \\
a_{CP}(\pi^{0}K^{-}) & \simeq-2\xi\frac{T}{P}\left[\sin(\delta_{T}-\delta_{P})+\frac{C}{T}\sin(\delta_{C}-\delta_{P})-r_{D}\sin(\delta_{T}-\delta_{D})\right]\sin\gamma.\label{eq:}\end{align}
 Numerical calculations for the direct CP asymmetries are given in
fig.\ref{fig:cpPIK}, which shows that when the value of $P$ is fixed
by the QCD factorization default value there is little room for $D$
except for the unreasonable region $D\gg P$. The strongest
constraint comes from $a_{cp}(\pi^{+}K^{-})$, for
$\delta_{D}=30^{\circ}$ the allowed value of $r_{D}$ is very narrow
around $r_{D}\simeq0.7$. Large $r_{D}$ is only allowed for some
special settings such as $\delta_{D}=120^{\circ}$. The data of
$a_{cp}(\pi^{0}K^{-})$ imposes a similar constraint, for
$\delta_{D}=30^{\circ}$ the allowed $r_{D}$ is around 0.5. For other
values of the strong phase $\delta_{D}=30^{\circ}\sim120^{\circ}$
the allowed $r_{D}$ is even smaller around $0.3.$ For the other two
modes $\pi^{0}\bar{K}^{0}$ and $\pi^{-}\bar{K}^{0}$ the constraints
are much weaker due to the weakened or vanishing tree-penguin
interferences. The data only disfavor the value of $r_{D}\sim1$. The
correlations among the four modes can be clearly seen from
fig.\ref{fig:cpPIK}. The charming penguin contribution in
$a_{cp}(\pi^{+}K^{-})$ is opposite to those in
$a_{cp}(\pi^{0}\bar{K}^{0})$ and $a_{cp}(\pi^{-}\bar{K}^{0})$ but
similar to $a(\pi^{0}K^{-})$. For small $\delta_{D}$ and
$r_{D}\sim1$, $a_{cp}(\pi^{+}K^{-})$ and $a(\pi^{0}K^{-})$ reach
their minimum, while $a_{cp}(\pi^{0}\bar{K}^{0})$ and
$a_{cp}(\pi^{-}\bar{K}^{0})$ close to their maximum.

Note that the charming penguin in $\Delta S=1$ modes are nearly CP
conserving, the time-dependent CP asymmetry for $\pi^{0}K_{S}$ remain
unchanged for small $r_{D}$ \begin{align}
S(\pi^{0}K_{S}) & \simeq\sin2\beta+2\xi\frac{C}{P}\cos(\delta_{C}-\delta_{P})\cos2\beta\sin\gamma\label{eq:}\end{align}
 This is due to the fact that the charming penguin contribution cancels
in the ratio $\bar{\mathcal{A}}/\mathcal{A}$ at the leading order.

\subsection{$KK$ modes}

The decay amplitudes for $KK$ modes are given by \begin{align}
\mathcal{\bar{\mathcal{A}}}(K^{+}K^{-}) & =-\lambda_{u}(E+P_{A})\nonumber \\
\mathcal{\bar{\mathcal{A}}}(K^{0}\bar{K}^{0}) & =-\lambda_{u}(P_{tu}-\frac{1}{3}P_{EW}^{C})-\lambda_{c}(P-D-\frac{1}{3}P_{EW}^{C})\nonumber \\
\mathcal{\bar{\mathcal{A}}}(K^{-}\bar{K}^{0}) & =\lambda_{u}(A-P_{tu}+\frac{1}{3}P_{EW}^{C})-\lambda_{c}(P-D-\frac{1}{3}P_{EW}^{C})\label{eq:}\end{align}
 The $K^{0}\bar{K}^{0}$ and $K^{-}\bar{K}^{0}$ are pure $\Delta S=0$
penguin modes. The charming penguin contribution is similar to that
of $\pi^{-}\bar{K}^{0}$ except for the CKM factors. From the latest
data, the QCD penguin amplitudes can be extracted and found consistent
with that from $\pi^{-}\bar{K}^{0}$. Note that in the SU(3) limit,
the direct CP asymmetry of $K^{0}\bar{K}^{0}$ is directly linked
to the $\pi^{-}\bar{K}^{0}$ \begin{align}
a_{CP}(K^{0}\bar{K}^{0}) & \simeq-2\frac{\omega r_{D}}{\omega^{2}-2\omega\cos\gamma+1}\sin(\delta_{D}-\delta_{P})\sin\gamma\nonumber \\
 & \simeq-\frac{\omega r_{D}}{\xi(\omega^{2}-2\omega\cos\gamma+1)}a_{CP}(\pi^{-}\bar{K}^{0})\label{eq:}\end{align}
Note that the asymmetry is enhanced by a factor $1/\xi\simeq50$.
From the current $1\sigma$ bound
$a_{CP}(\pi^{-}\bar{K}^{0})=0.009\pm0.025$,
$a_{CP}(K^{0}\bar{K}^{0})$ can easily reach to $\mathcal{O}(-0.5)$
for $r_{D}\sim0.5$. A stronger $r_{D}$ dependence can be found in
the time-dependent CP asymmetry of $K_{S}K_{S}$. For small $r_{D}$
the quantity $S(K_{S}K_{S})$ is nonzero \begin{eqnarray}
S(K_{S}K_{S}) & \simeq &
\frac{2\omega}{\sqrt{\omega^{2}-2\omega\cos\gamma+1}}r_{D}\cos(\delta_{D}-\delta_{P})\sin\beta.\end{eqnarray}
 For $r_{D}\sim0.5$ and $\delta_{D}-\delta_{P}\sim0^{\circ}$, the
CP asymmetry can reach to be $S(K_{S}K_{S})\sim$0.5. %
\begin{figure}
\includegraphics[width=0.45\textwidth]{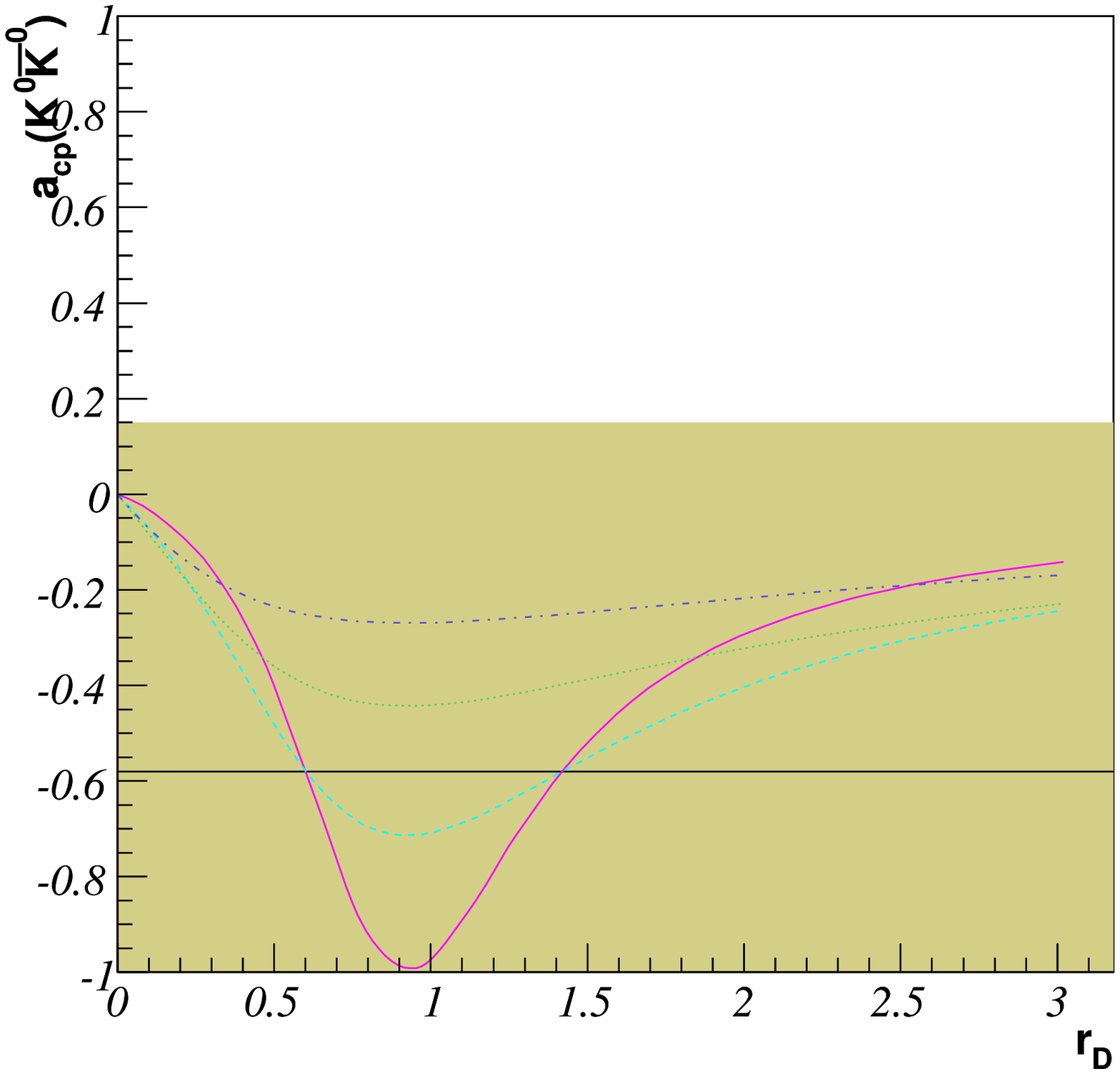} \includegraphics[width=0.45\textwidth]{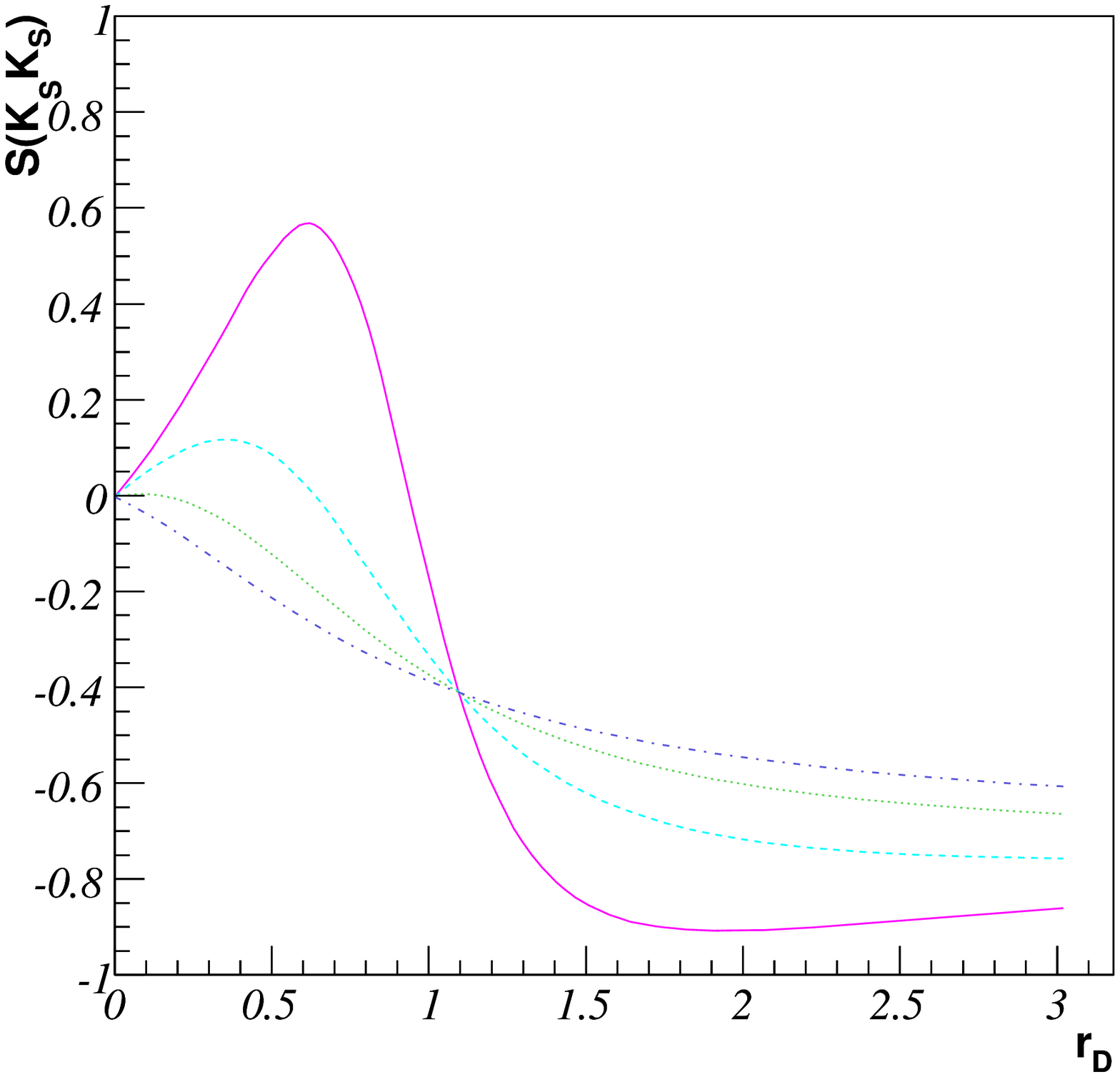}

\caption{\label{fig:cpPIPI}$a_{cp}(K^{0}\bar{K}^{0})$ and $S(K_{S}K_{S})$
as function of $r_{D}$. Four curves corresponds to the strong phase
$\delta_{D}=30^{\circ}$(solid), $60^{\circ}$(dashed), $90^{\circ}$(dotted)
and $120^{\circ}$(dot-dashed) respectively. Other parameters are
default in QCD factorization estimations. }

\end{figure}

\section{\label{sec:Constrining}Constraining charming penguins from global
fit}

Let us go a step further for a model independent determination of
charming penguin. Since the charming penguin and QCD penguin always
come together, distinguishing the two relies on their different
interference with tree type diagrams $T,C$ and the different
contributions for $\Delta S=0$ and $1$ modes. It also depends
heavily on the precision of the experimental data on CP asymmetries.
To isolate $SU(3)$ breaking and possible new physics effects, we
shall proceed in two steps: i) Fit only to $\Delta S=0$ modes
including 11 available data points in $\pi\pi$ and $KK$ in $SU(3)$
limit, which is a minimal set to determine the seven parameters $T$,
$C$, $\delta_{C}$, $P$, $\delta_{P}$, $D$ and $\delta_{D}$. The data
for $K^{+}K^{-}$ is excluded, because it constrains only the
annihilation diagram $E$. The subleading $P_{EW}$ is fixed to $T+C$
through the SM relation from the isospin analysis
\cite{Neubert:1998re,Gronau:2000pk,Wu:2002nz}. The advantages to use
this data set are that the $\Delta S=0$, $b\to d$ modes are expected
to have less $SU(3)$ breaking and less affected by possible new
physics. The main disadvantage is that the accuracy is limited by
fewer data points. ii) Fits to both $\Delta S=0$ and $1$ modes which
includes 19 available data points in $\pi\pi,$ $KK$ and $\pi K$
modes. Using the approximate $SU(3)$ symmetry, the fit accuracy is
greatly improved. The stability of the result is checked by fit with
different $SU(3)$ breaking schemes. As already mentioned, the
potential puzzles in $B\to\pi\pi$ and $\pi K$ modes can be divided
into hadronic dynamics related and new physics related ones. Since
the current data show a significant reduction of the $\pi K$ puzzle
in decay rates, the implication of new physics beyond the SM is
mostly related to the low $S(\pi^{0}K_{S})$ which remains to be
confirmed by future experiments. We shall exclude this data point in
the fits because they have little effects in determining the
charming penguin and shall discuss it separately.

There already exists a number of global fits to charmless $B$ decays
using flavor diagrammatic methods \cite{Wu:2000rb,Zhou:2000hg,Chiang:2004nm,Wu:2004xx,Wu:2005hi}
and flavor $SU(3)$ invariant amplitudes \cite{he:2000ys,He:2004ck,Fu:2002nr},
which focus on using the data as an independent determination of the
weak phases in the unitarity triangle. A recent analysis \cite{Chiang:2006ih}
shows an essential agreement with the global CKM fit results on the
profile of UT \cite{Charles:2004jd}. Since the purpose of the present
work focuses on the charming penguin induced FSI, we simply take the
values of $V_{ub}$ , $V_{cb}$ and the weak phase from the global
CKM fit given in Eq.(\ref{eq:Vub-Vcb-input}) and (\ref{eq:gamma-input})
as inputs to further reduce the uncertainties. In fact, it was shown
in ref.\cite{Wu:2005hi} that the resulting weak phase $\gamma$
from a model independent global fit is consistent with the standard
model and insensitive to the various cases, such as the new physics
effects in electroweak penguin sector, the SU(3) flavor symmetry breaking
effects in strong phase and the charming penguin effects. As a convention,
all the $Br$s are in units of $10^{-6}$ and the phase angles are
in gradient and arranged in the range of $(-\pi,+\pi)$.

\begin{table}
\input{data.tex}

\caption{\label{cap:the-latest-data}Experimental data for charmless $\Delta S=0$
and $\Delta S=1$ $B$ decay modes}

\end{table}

\subsection{Fit to $\pi\pi$ and $KK$ modes}

The fit to $\pi\pi$ and $KK$ decay modes are summarized in Fit.1a(b)
in Tab.\ref{cap:Tab-1}. For comparison purpose, we give in Fit.1a
a determination of the leading diagrams without charming penguin amplitude.
The result is characterized by a large $C/T$ and also a slightly
large $P/T$ compared with short distance QCD factorization description
\begin{eqnarray}
\frac{C}{T}=0.63\pm0.09 & , & \ \frac{P}{T}=0.19\pm0.02\ (Fit.1a).\end{eqnarray}
 In fit.1b the charming penguin contribution is switched on. One
sees that there is a significant reduction of $C/T\sim0.35$ from
the best fitted central values while $P/T$ is further enhanced. The
size of $D$ is found smaller than that of $P$ \begin{eqnarray}
\frac{C}{T}=0.35\pm0.16 & , & \ \frac{P}{T}=0.36\pm0.19,\
r_{D}=0.63\pm0.62\ (Fit.1b).\end{eqnarray}
 The best fits favor a constructive interference between $C$ and
$P$ which is driven by the large decay rate of $\pi^{0}\pi^{0}$.
The interference between $P$ and $D$ is largely destructive, which
compensates the growth of $P$. Due to the limited degree of freedom,
the inclusion of charming penguin leads to large uncertainties in
all the fitted parameters. The $\chi_{min}^{2}$ curve for $D$ given
in Fig.\ref{fig:chisq} shows a rather flat dependence of $\chi_{min}^{2}$,
which sets a 1$\sigma$ upper bound of $r_{D}<1.25$. For a meaningful
determination of $D$, more precise data for penguin dominant $\pi K$
and $KK$ modes are needed. The prediction for the yet to be measured
modes are \begin{eqnarray}
a_{cp}(\pi^{0}\pi^{0})=0.29\pm0.48 & , & S(\pi^{0}\pi^{0})=0.77\pm0.58,\nonumber \\
a_{cp}(K^{0}\bar{K}^{0})=0.08\pm0.52 & , & S(K_{S}K_{S})=0.93\pm0.44.\end{eqnarray}
 The predicted $a_{cp}(K^{0}\bar{K}^{0})$ is small but $S(K_{S}K_{S})$
is very large, which follows from the best fitted $\delta_{D}-\delta_{P}\simeq0$
and a large $r_{D}=0.62.$ The uncertainties in the predictions are
also large.

\begin{table}
\input{bstFitTab1.tex}

\caption{\label{cap:Tab-1}Hadronic parameters determined from global fit to
the data. Fit.1a: fit to $\pi\pi$ and $KK$ modes in SU(3) limit
without charming penguin. Fit1b: the same as Fit.1a with charming
penguin included. Fit2a: fit to $\pi\pi$, $\pi K$ and $KK$ modes
in SU(3) limit without charming penguin. Fit2b: the same as Fit2a
with charming penguin included.}

\end{table}

\subsection{Fit to $\pi\pi$, $KK$ and $\pi K$ modes}

A stronger constraint can be obtained by including the $\pi K$ modes
using flavor $SU(3)$ symmetry. The $\Delta S=1$ modes are penguin
dominant, which constrains mostly the combination $P-D$, and also
their relative phase from direct CP asymmetries. Although the
current $\pi K$ data only established the direct CP asymmetry in
$\pi^{+}K^{-}$, nontrivial bounds for other modes are already
obtained. The fit with $\pi K$ in $SU(3)$ limit is listed in
Tab.\ref{cap:Tab-1} (Fit 2). In the case of no charming penguin
(Fit.2a), one sees an even larger $C/T\simeq0.7$ which is known to
be driven by the $\pi K$ CP puzzle, and the ratio $P/T$ slightly
reduced \begin{eqnarray} \frac{C}{T}=0.72\pm0.08 & , & \
\frac{P}{T}=0.19\pm0.01\ (Fit.2a).\end{eqnarray}
 Note that the inclusion of $\pi K$ modes leads to a significant
reduction of the uncertainty.

The fit including the charming penguin is given in Fit.2b. Unlike
the previous fits, when the $\pi K$ modes are included, the ratio
$C/T$ remains large. The inclusion of charming penguin only leads
to a slight reduction for $C/T$ from $\sim$0.72 to $\sim0.58$,
due to the lower value of $r_{D}$ around $0.43$ with an improved
precision \begin{eqnarray}
\frac{C}{T}=0.58\pm0.14 & , & \
\frac{P}{T}=0.28\pm0.17,\ r_{D}=0.43_{-0.43}^{+0.65}\ (Fit.2b).\end{eqnarray}
 Compared with Fit.1a, the absolute size of $D$ is reduced from 0.18
to 0.078. With the improved precision, the fit result indicates stronger
constraint on $r_{D}$. This can be seen from the $\chi^{2}$ curve
in Fig.6. The prediction for the CP asymmetries in $\pi^{0}\pi^{0}$
and $K^{0}\bar{K}^{0}$ modes are \begin{eqnarray}
a_{cp}(\pi^{0}\pi^{0})=0.53\pm0.15 & , & S(\pi^{0}\pi^{0})=0.73\pm0.16,\nonumber \\
a_{cp}(K^{0}\bar{K}^{0})=-0.04\pm0.22 & , & S(K_{S}K_{S})=0.46\pm0.44.\end{eqnarray}
 The reduction of $S(K_{S}K_{S})$ is also related to the reduced
$r_{D}$.

To check the SU(3) breaking effects, in Tab.\ref{cap:Tab-2} we list
the fit results for two SU(3) breaking scheme: one is for SU(3) breaking
in $T$ diagrams only (Fit.3), the other one is for SU(3) breaking
for both $T$ and $C$ (Fit.4). The SU(3) breaking factor is set to
$f_{K}/f_{\pi}=1.22$ for $\Delta S=1$ modes. The obtained results
show the value of $r_{D}$ is quite stable

\begin{figure}
\includegraphics[width=0.7\textwidth]{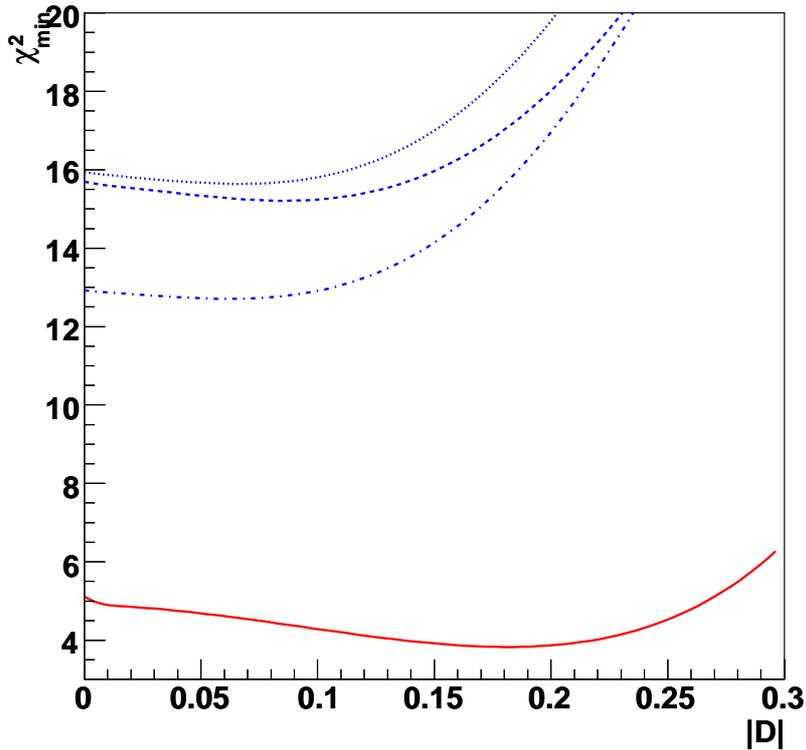}

\caption{\label{fig:chisq}$\chi^{2}$ as functions of $D$. The solid, dashed,
dotted and dot-dashed curves correspond to Fit.1b, 2b,3b and 4b respectively.}

\end{figure}

\begin{table}
\input{bstFitTab2.tex}

\caption{\label{cap:Tab-2}Hadronic parameters determined from global fit to
$\pi\pi$, $\pi K$ and $KK$. Fit.3a: fit without charming penguin.
A SU(3) breaking factor $f_{K}/f_{\pi}$is associated to tree diagrams.
Fit3b: the same as Fit.3a with charming penguin included. Fit4a: fit
without charming penguin. The SU(3) breaking factor $f_{K}/f_{\pi}$is
associated to both tree and color-suppressed tree diagrams. Fit4b:
the same as Fit4a with charming penguin included.}

\end{table}

\begin{eqnarray}
\frac{C}{T}=0.58\pm0.16 & , & \ \frac{P}{T}=0.29\pm0.12,\ r_{D}={0.34}_{-0.34}^{+0.39}\ (Fit.3b)\nonumber \\
\frac{C}{T}=0.57\pm0.17 & , & \ \frac{P}{T}=0.29\pm0.13,\
r_{D}={0.31}_{-0.31}^{+0.47}\ (Fit.4b)\end{eqnarray}
 The corresponding $\chi^{2}$ curves are shown in Fig.\ref{fig:chisq}.
The $SU(3)$ breaking scheme in Fit.4b gains the lowest $\chi^{2}$,
in a good agreement with previous analysis \cite{Chiang:2006ih}
on SU(3) breaking. The prediction from Fit4.b are given by%
{}{}{} \begin{eqnarray}
a_{cp}(\pi^{0}\pi^{0})=0.54\pm0.25 & , & S(\pi^{0}\pi^{0})=0.74\pm0.22\nonumber \\
a_{cp}(K^{0}\bar{K}^{0})=-0.02\pm0.22 & , & S(K_{S}K_{S})=0.38\pm0.19\end{eqnarray}
Thus all the Fit.1-4 favor a small $a_{cp}(K^{0}\bar{K}^{0})$ compatible
with zero as a consequence of $\delta_{D}\simeq\delta_{P}$ but a
positive $S(K_{S}K_{S})=0.3\sim0.4$. This kind of pattern is unique
for the charming penguin contribution, which can be used to distinguish
it from other contributions such as possible new physics from electroweak
penguin sector%
{}{}{}. Note that the current data of $S(K_{S}K_{S})$ are not yet
conclusive. The Babar and Belle collaborations report \cite{Aubert:2006gm,K.Abe:2007xd}
\begin{eqnarray}
S(K_{S}K_{S})=-1.28_{-0.73-0.16}^{+0.80+0.11} & , & C(K_{S}K_{S})=-0.40\pm0.41\pm0.06(Babar)\nonumber \\
S(K_{S}K_{S})=-0.38\pm0.77\pm0.08 & , & C(K_{S}K_{S})=+0.38\pm0.38\pm0.05(Belle)\end{eqnarray}
 The Babar result for $S(K_{S}K_{S})$ favors a value outside physical
region and has different sign for $C(K_{S}K_{S})$. 
In the Fit.2-4,
including the two free parameters $D$ and $\delta_{D}$ only leads to
slight reduction of the $\chi_{min}^{2}$ from 4.5 to 3.9 for Fit.1b
(from 12.2 to 11.8 for Fit.4b). Thus the charming penguin does not
play an important role to improve the agreement with the data. The
best fitted $C/T$ remains large around $0.6$. The charming penguin
can not play a significant role in reducing the $\pi\pi$ puzzle.

We have checked the influence of the measurement of $S(\pi^{0}K_{S})$.
Including this piece of data leads to a big increase of the $\chi^{2}$
but all the best fitted parameters remain unchanged. For instance,
in the SU(3) breaking scheme of Fit.4b, we get $\chi^{2}=17.7$, $C/T=0.58\pm0.12$
and $r_{D}=0.34_{-0.34}^{+0.36}$. As mentioned before, a low $S(\pi^{0}K_{S})$
can hardly be accommodated within the SM, and can be a signal of new
physics. A possibility is that $P_{EW}$ carries a large $CP$ phase
\cite{Yoshikawa:2003hb,Buras:2003dj,Buras:2004th,Wu:2006ur}. Some 
strategies for extracting new physics parameters are discussed in
Refs.\cite{London:2004ej,London:2004uj}. 
On the other hand, the $\pi\pi$ and $\pi K$ CP puzzle are more relevant
to the low energy hadronic dynamics and should be investigated separately.

\section{\label{sec:conclusions}discussions and conclusions}

An equivalent way to see the charming penguin effects is to take $P_{tu}$
as a free parameter not equal to $P$ and avoiding introducing the
amplitude $D$. A global fit including $P_{tu}$ was done a few years
ago which favored a large difference between $P_{tu}$ and $P$, hence
a large charming penguin was implied \cite{Chiang:2004nm}. However,
the data have been significantly  changed over the years. The main changes
in the data include i) a reduced $\pi\pi$ puzzle from $R_{\pi\pi}=0.83\pm0.23$
to the current value of $0.51\pm0.08$ ; ii) a reduced but more precise
value of $S(\pi^{+}\pi^{-})$ from $-0.70\pm0.30$ to $-0.61\pm0.08$;
iii) a more precise $a_{cp}(\pi^{+}K^{-})$ from $-0.09\pm0.03$ to
$-0.097\pm0.01$ and $a_{cp}(\pi^{0}K^{-})$ from $0.00\pm0.12$ to
$0.05\pm0.025$. The updated data are moving towards a much stronger
constraints on the charming penguin. Our present conclusion is therefore
different from the previous one. %
{}{}{}{}

In conclusion, we have found strong constraints to the charming
penguins from its correlated contributions to $B\to\pi\pi,\pi K$ and
$KK$ decay modes. These correlations are illustrated by adding the
charming penguin amplitudes to these decay modes while assuming that
other hadronic amplitudes are short-distance dominated. The charming
penguin contribution has negative correlations between
$Br(\pi^{+}\pi^{-})$ and $Br(\pi^{0}\pi^{0})$,
$a_{cp}(\pi^{+}\pi^{-})$ and $a_{cp}(\pi^{0}\pi^{0})$,
$a_{cp}(\pi^{+}K^{-})$ and $a_{cp}(\pi^{0}\bar{K}^{0})$
respectively. Positive correlations are found between
$a_{cp}(\pi^{+}K^{-})$ and $a_{cp}(\pi^{-}\bar{K}^{0})$,
$a_{cp}(\pi^{0}\bar{K}^{0})$ and $a_{cp}(\pi^{-}\bar{K}^{0})$. These
correlations are unique nature of the charming penguin, and can be
used to distinguish its contribution from the others. Using the
latest data and assuming the approximate flavor $SU(3)$ symmetry,
the size of charming penguin is constrained from a global fit. The
results favor a small \[ r_{D}\simeq0.3-0.4\mbox{ and
}\delta_{D}\simeq\delta_{P}.\]
 which makes it unlikely as a solution to the $\pi\pi$ puzzle. The
color-suppressed tree amplitude remains large $C/T\simeq0.6$. The
time-dependent CP asymmetries in $\pi^{0}\pi^{0}$ and $K_{S}K_{S}$
modes are highly sensitive to the charming penguin. We have found
that charming penguin leads to a sizable $S(K_{S}K_{S})\approx0.3$
while keep $a_{cp}(K^{0}\bar{K}^{0})$  compatible with zero.

\begin{acknowledgments}
We are grateful to G.Hiller for helpful discussions and early involvement
of the present work. This work is supported in part by the National
Science Foundation of China (NSFC) under the grant 10475105, 10491306,
and the key Project of Chinese Academy of Sciences (CAS).

\bibliography{/home/zhou/reflist/reflist,/home/zhou/reflist/temp,/home/zhou/reflist/misc}

\end{acknowledgments}

\end{document}

%% file: data.tex

\begin{ruledtabular} \begin{tabular}{llll}
modes  & $Br$($\times10^{-6}$)  & $a_{CP}$  & $S$\tabularnewline
\hline 
$\pi^{+}\pi^{-}$  & $5.16\pm0.22$  & $0.38\pm0.07$  & $-0.61\pm0.08$\tabularnewline
\hline 
$\pi^{0}\pi^{0}$  & $1.31\pm0.21$  & $0.48_{-0.31}^{+0.32}$  & \tabularnewline
\hline 
$\pi^{-}\pi^{0}$  & $5.59_{-0.40}^{+0.41}$  & $0.06\pm0.05$  & \tabularnewline
\hline 
$\pi^{+}K^{-}$  & $19.4\pm0.6$  & $-0.097\pm0.012$  & \tabularnewline
\hline 
$\pi^{0}\bar{K}^{0}$($K_{S}$)  & $9.9\pm0.6$  & $-0.14\pm0.11$  & $(0.38\pm0.19)$\tabularnewline
\hline 
$\pi^{-}\bar{K}^{0}$  & $23.1\pm1.0$  & $0.009\pm0.025$  & \tabularnewline
\hline 
$\pi^{0}K^{-}$  & $12.9\pm0.6$  & $0.050\pm0.025$  & \tabularnewline
\hline 
$K^{+}K^{-}$  & $0.07\pm0.12$  &  & \tabularnewline
\hline 
$K^{0}\bar{K}^{0}$  & $0.96_{-0.19}^{+0.21}$  & $-0.58_{-0.66}^{+0.73}$  & \tabularnewline
\hline 
$K^{-}\bar{K}^{0}$  & $1.36_{-0.27}^{+0.29}$  & $0.12_{-0.18}^{+0.17}$  & \tabularnewline
\end{tabular}\end{ruledtabular} \label{data} 

%% file: bstFitTab1.tex
\begin{ruledtabular} \begin{tabular}{lllll}
parameter  & Fit 1a  & Fit 1b  & Fit 2a  & Fit 2b\tabularnewline
\hline 
$T$  & $0.652\pm0.036$  & $0.799_{-0.151}^{+0.090}$  & $0.649\pm0.035$  & $0.720\pm0.111$ \tabularnewline
$C$  & $0.408\pm0.052$  & $0.282_{-0.071}^{+0.115}$  & $0.467\pm0.044$  & $0.416_{-0.067}^{+0.082}$ \tabularnewline
$\delta_{C}$  & $-0.774_{-0.212}^{+0.250}$  & $-0.983_{-0.365}^{+0.408}$  & $-1.096\pm0.132$  & $-1.189\pm0.229$ \tabularnewline
$P$  & $0.124\pm0.010$  & $0.290_{-0.141}^{+0.065}$  & $0.124\pm0.006$  & $0.201\pm0.119$ \tabularnewline
$\delta_{P}$  & $-0.450_{-0.111}^{+0.097}$  & $-0.465\pm0.115$  & $-0.429\pm0.052$  & $-0.362\pm0.116$ \tabularnewline
$P_{EW}$  & $0.013\pm0.001$  & $0.013\pm0.001$  & $0.013\pm0.001$  & $0.013\pm0.001$ \tabularnewline
$\delta_{P_{EW}}$  & $-0.294\pm0.103$  & $-0.241\pm0.102$  & $-0.448\pm0.066$  & $-0.416\pm0.078$ \tabularnewline
$D$  & $0(fixed)$  & $0.182_{-0.156}^{+0.080}$  & $0(fixed)$  & $0.078\pm0.119$ \tabularnewline
$\delta_{D}$  & $0(fixed)$  & $-0.497_{-0.512}^{+0.172}$  & $0(fixed)$  & $-0.307\pm0.294$ \tabularnewline
$\chi^{2}/dof$  & 4.4/6  & 3.8/4  & 15.7/12  & 15.2/10 \tabularnewline
\end{tabular}\end{ruledtabular}

%% file: bstFitTab2.tex
\begin{ruledtabular} \begin{tabular}{lllll}
parameter  & Fit 3a  & Fit 3b  & Fit 4a  & Fit 4b\tabularnewline
\hline 
$T$  & $0.651\pm0.035$  & $0.713\pm0.092$  & $0.650\pm0.035$  & $0.708_{-0.087}^{+0.077}$ \tabularnewline
$C$  & $0.462\pm0.043$  & $0.419_{-0.063}^{+0.079}$  & $0.457\pm0.043$  & $0.415_{-0.064}^{+0.081}$ \tabularnewline
$\delta_{C}$  & $-1.080\pm0.135$  & $-1.163\pm0.199$  & $-1.052\pm0.127$  & $-1.121\pm0.188$ \tabularnewline
$P$  & $0.124\pm0.007$  & $0.191\pm0.095$  & $0.124\pm0.006$  & $0.186_{-0.067}^{+0.076}$ \tabularnewline
$\delta_{P}$  & $-0.362\pm0.043$  & $-0.311\pm0.116$  & $-0.372\pm0.044$  & $-0.328\pm0.121$ \tabularnewline
$P_{EW}$  & $0.013\pm0.001$  & $0.013\pm0.001$  & $0.013\pm0.001$  & $0.013\pm0.001$ \tabularnewline
$\delta_{P_{EW}}$  & $-0.439\pm0.066$  & $-0.413_{-0.081}^{+0.073}$  & $-0.425\pm0.061$  & $-0.398_{-0.081}^{+0.069}$ \tabularnewline
$D$  & $0(fixed)$  & $0.067\pm0.095$  & $0(fixed)$  & $0.062\pm0.100$ \tabularnewline
$\delta_{D}$  & $0(fixed)$  & $-0.260\pm0.363$  & $0(fixed)$  & $-0.295\pm0.388$ \tabularnewline
$\chi^{2}/dof$  & 15.9/12  & 15.6/10  & 12.9/12  & 12.7/10 \tabularnewline
\end{tabular}\end{ruledtabular}